\documentstyle{article}
\begin{document}
\LARGE
\begin{center}
\vspace*{0.3in} \bf No-boundary Codimension-two Braneworld
\vspace*{0.6in} \normalsize \large

\rm Zhong Chao Wu

Dept. of Physics

Zhejiang University of Technology

Hangzhou 310032, China

\vspace*{0.4in} \large \bf Abstract
\end{center}
\vspace*{.1in} \rm \normalsize \vspace*{0.1in}

The quantum creation probability and entropy of a 2-codimensional
braneworld are calculated in the framework of no-boundary
universe. The entropy can take an arbitrarily large value as
the brane tensions increase, in violation of the conjectured
``$N$-bound" in quantum gravity, even for a 4-dimensional ordinary universe.

 \vspace*{0.3in}

PACS number(s): 04.50.+h; 98.80.Cq; 11.10.Kk; 04.65.+c


\vspace*{0.5in}

e-mail: zcwu@zjut.edu.cn,

\pagebreak

In recent years, the so-called braneworld has attracted a lot of
attention. Here the Standard Model fields are confined in the
4-dimensional brane, which is embedded in a higher dimensional
bulk space. The extra dimensions can only be perceived through
gravitational interactions. They need not necessarily be small and
may even be on the scale of millimeters. This concept provides a
possible solution to the hierarchy problem; that is the large
difference between the Planck and electroweak scales [1]. Randall
and Sundrum [2] showed that the 4-dimensional Newton gravity can
be reproduced in the brane even in the presence of an infinitely
large extra dimension. In the $RS$ model the bulk spacetime is
described by a 5-dimensional Anti-de Sitter spacetime.

The interest of brane cosmology was inspired by string theory.
Horava and Witten [3] suggested an 11-dimensional spacetime model.
The eleventh dimension is compactified on an orbifold with $Z_2$
symmetry, while 6 dimensions can be compactified on a Calabi-Yau
manifold in the usual way.

On the other hand, it is believed that the quantum state of the
universe is defined by the ground state in no-boundary universe
[4]. The wave function of the universe is expressed as the path
integral over all Euclidean metrics and matter fields on them. The
contribution of an instanton solution dominates the path integral.
Therefore, at the $WKB$ level, the Lorentzian universe can be
obtained via analytic continuation from the instanton. Even though
the no-boundary proposal works only for a closed universe, one can
still analytically continue a complex instanton to obtain an open
universe, at the same $WKB$ level of approximation [5].

Some work have been done on the quantum creation of braneworlds
[6]. Some of this work has been based on the $RS$ model,
unfortunately, there does not exist a compact seed instanton for
it. Although the most effort has been paid to the 1-codimensional
braneworld, higher codimensional models are attracting interest
recently [7]. It is hoped that some of 2-codimensional models
provide a distinctive step toward understanding the smallness of
the cosmological constant. In this paper we shall discuss a
creation scenario of 2-codimensional braneworld.

For generality and simplicity, the bulk dimension is not specified
here. The Euclidean action for a $n$-dimensional bulk spacetime
$M$ with a brane $B$ is
\begin{equation}
I = -\frac{1}{16\pi }\int_M (R - 2\Lambda) -\frac{1}{8\pi}
\int_{\partial M} K + \frac{1}{8\pi}\int_B \sigma,
\end{equation}
where the $n-$dimensional Newton constant is set to be $1$,
$\Lambda$ is the bulk cosmological constant, the scalar curvature
$R$ and the trace of the second fundamental form $K$ are defined
with respect to the bulk. The boundary $\partial M$ may be
degenerate and then $K$ can take distribution form. This is the
case for the boundary associated with the 2-codimensional branes.
We have assumed that there is no matter content in the bulk except
for a cosmological constant $\Lambda$, the only energy-momentum
tensor of the brane is represented by the tension $\sigma$.

We first consider the Schwarzschild-like instanton
\begin{equation}
ds^2 = f(r)d\psi^2 + f(r)^{-1}dr^2 + r^2 d\Omega^2_{n-2},
\end{equation}
where
\begin{equation}
f(r)= k - \frac{2m}{r^{n-3}} - \frac{2\Lambda r^2}{(n-1)(n-2)},
\end{equation}
where $m$ is a parameter.
$d\Omega^2_{n-2}$ is the metric of an unit $n-2-$dimensional Einstein
manifold, called the base manifold. If it is restricted to be
maximally symmetric, then it must be either a $n-2-$dimensional
sphere, a plane or a hyperboloid labelled by $k = 1,0,-1$,
respectively. The hyperboloid and plane can be compactified using
their discrete isometry subgroup. The sphere can also be replaced
by its lens space.

In general, there are $n-1$ horizons $r_i$ associated with the
zeros of $f(r)$. The surface gravity at the horizon $r_i$ is
\begin{equation}
\kappa_i =  \frac{1}{2}\left |\frac{df(r)}{dr}\right |_{r =r_i}.
\end{equation}
One can construct a compact manifold by periodically identifying
$\psi$ coordinate between two neighboring horizons $r_l \leq r
\leq r_k $, with $f(r) \geq 0$. Since the motivation of this article is
to discuss the braneworld, there is no restriction for the signature of the parameter $m$.
Therefore, for $k =0, -1$, in order for the two horizons to exist,
there is no restriction for the signature of the cosmological constant either.

If one chooses the period $\beta_l
= 2\pi \kappa^{-1}_l$, then the conical singularity at the horizon
$r_l$ can be regularized. However, in general the two surface
gravities are distinct, therefore one cannot construct a regular
instanton.

One can keep the identified coordinate $\psi$ of metric (2)
intact and analytically continue the base space $d\Omega^2_{n-2}$.
In this case, the base space becomes a $n-2$-dimensional de
Sitter, Minkowski, Anti-de Sitter or their respective lens
spacetime. The analytic continuation from the sphere $S_{n-2}$ of
the base space to the $n-2-$dimensional de Sitter space is well
known [4]. From the hyperboloid $H_{n-2}$ of the base space
\begin{equation}
ds^2= d\theta^2 +  \sinh^2 \theta dS^2_{n-3}=d\theta^2 +  \sinh^2
\theta (d\kappa^2 + \sin^2 \kappa dS^2_{n-4})
\end{equation}
one can obtain the $n-2-$dimensional Anti-de Sitter space
\begin{equation}
ds^2 = - dt^2 + \cos^2t(d\phi^2 + \sinh^2 \phi dS^2_{n-4})=- dt^2
+ \cos^2t dH^2_{n-3} ,
\end{equation}
by the analytic continuation $t = i \theta + \pi/2$ and $\phi = i
\kappa$, where $dS^2_d$ and $dH^2_d$ represent the metrics of unit
$S_d$ and $H_d$, respectively.

This technique is called double analytic continuation, which was
first introduced by Witten to study the instability of the
Kaluza-Klein vacuum [8]. The obtained Lorentzian manifolds, dubbed
``bubbles of nothing", can be considered as time dependent
backgrounds for string theory [9][10]. It is noted that the
backgrounds must be regular and no singularity is allowed.
Therefore, in the literature, the technique of double continuation
is applied only to a black hole with asymptotically flat or
Anti-de Sitter background.

The motivation of this paper is different. We are going to use the
same seed instanton (2) for braneworld creation. At the $WKB$
level, the braneworld can be obtained by the double continuation.
We identify the two conical singularities at the horizons as the
2-codimensional  branes embedded in the bulk. The brane in which
we are living is a $n-2-$dimensional de Sitter, Minkowski or
Anti-de Sitter spacetime.

The  Einstein equation should be satisfied everywhere, the brane
tension $\sigma$ is the source of the conical singularity. Since
the field configuration is invariant with respect to the Lorentz
boosts or rotation in the local inertial frame of the base
manifold, the tension should behave as a cosmological constant in
the base manifold. That is, the pressure $p$ is the negative of
the energy density which is $\sigma$. This fact has been accounted
in the brane part of the action (1).

Since the conical singularities are allowed for the constructed
Euclidean manifold, the identification period $\beta$ can be
relaxed from its relation with the surface gravity. The bulk
equator of the quantum transition topologically is the product of
the space $M_2$ associated with coordinates $(\psi, r)$ and an
equator of the $n-2$-dimensional base manifold.

We use $M_l$ and $M_k$ to denote the infinitesimal neighborhoods
of the two horizons with the boundary of a constant coordinate
$r$, and we use $M^\prime$ to represent $M$ minus $M_l$ and $M_k$.
The Euclidean action can be rewritten as
\begin{equation}
I = I_l + I_k + \int_{M^\prime} (\pi^{ij}\dot{h}_{ij} - NH_0 -N_i
H^i)d^nx + \frac{1}{8\pi}\int_B \sigma,
\end{equation}
where the actions $I_l$ and $I_k$ are the gravitational actions
for $M_l$ and $M_k$, respectively. The action of $M^\prime$ has
been recast into the canonical form. $N$ and $N_i$ are the lapse
function and shift vector, $h_{ij}$ and $\pi^{ij}$ are the
$n-1$-metric and the conjugate momenta respectively, $H_0$ and
$H^i$ are the Einstein and momentum constraints, and the dot
denotes the derivative with respect to the Killing coordinate
$\psi$. The manifold satisfies the Einstein equation, and all time
derivatives vanish due to the $U(1)$ isometry. Therefore, the
integral over $M^\prime$ is zero.

Now the action $I_l$ or $I_k$ can be written
\begin{equation}
I_i =- \frac{1}{16 \pi} \int_{M_i} ({}^nR - 2 \Lambda) -
\frac{1}{8\pi} \int_{\partial M_i }{}^{n-1}K  \;\; (i = l, k),
\end{equation}
where $^{n}R$ denotes the $n$-dimensional scalar curvature and
$^{n-1}K$ is the expansion rate of the boundary. In addition to
the action from the boundary of $M_i$, the action contribution of
the conical singularity can be considered as the degenerate
version of the second term. The conical singularity contribution
is termed as a deficit ``angle''. We assume $r_i$ to be a single zero
of $f(r)$, therefore the horizon does not recede into an internal
infinity.

One can apply the Gauss-Bonnet theorem to the 2-dimensional
$(\psi, r)$ section of $M_i$,
\begin{equation}
 \frac{1}{4 \pi} \int_{\hat{M}_i}{}^2R +
\frac{1}{2\pi} \int_{\partial \hat{M}_i }{}^1K  +
\frac{\delta_i}{2\pi} = \chi (i),
\end{equation}
where $\hat{M}_i$ is the projection of $M_i$ onto the
2-dimensional $(\psi, r)$ section, ${}^2R$ is the scalar curvature
on it, ${}^1K$ is the corresponding expansion rate, $\delta_i$ is
the deficit angle at the apex, and $\chi(i)$ is the Euler
characteristic of $\hat{M}_i$, which is $1$ here. Since the
expansion rate of the subspace $r^2 d\Omega^2_{n-2}$ goes to zero
at the horizon, ${}^{n-1}K$ and ${}^1K$ are equal. Comparing Eqs.
(8) and (9), one can see that as the circumference of the boundary
tends to zero, the action (8) becomes $-\chi(i) A_i/4$, where
$A_i= r^{n-2}_i \Omega_{n-2}$ is the $n-2-$dimensional surface
area of the horizon. It is noted that both the volume integral of
(8) and the first term of the left hand side of (9) vanish as the
boundary approaches the horizon.

Therefore, the total gravitational action is
\begin{equation}
I_{gravity} = -\frac{1}{4} (A_l + A_k).
\end{equation}
That is the negative of one quarter of the sum of the two horizon
areas. It is worth emphasizing that the horizon areas are the
volumes of the $n-2-$dimensional Euclidean brane manifolds.

In order to find the tension of the brane for the given $\beta$,
we first assume that the brane has 2-dimensional finite
``thickness", and then let the thickness approach zero later. One
can use the Gauss-Bonnet theorem again to the 2-dimensional cross
section of $\hat{M_i}$,
\begin{equation}
\frac{\alpha_i}{2\pi} \equiv \chi (i) -\frac{1}{2\pi}
\int_{\partial \hat{M_i}} {}^1K
 = \frac{1}{4\pi}\int_{\hat{M_i}} {}^2R,
\end{equation}
where the deficit angle $\alpha_i$ is redefined as the orientation
change of a parallel transport of a vector around $\partial
\hat{M_i}$ instead of that defined at the apex in equation (9). As
the thickness is reduced to zero, the limit of $\alpha_i$ becomes
$\delta_i$.

We assume the tension $\sigma_i$ is diluted to a tension density
$\epsilon_i (\psi, r)$ with a finite support ( or the thickness)
such that
\begin{equation}
\sigma_i = \int_{\hat{M_2}} \epsilon_i (\psi, r).
\end{equation}
Since the thickness will be infinitesimal, as long as the relation
of the deficit angle and the tension is concerned, the bulk
cosmological constant can be ignored. For the same consideration,
it is also assumed that the spacetime $\bar{M}_n$ near the brane $r_i$
can be thought as a direct product of the base manifold $M_{n-2}$
and the cross section $\bar{M}_2$
\begin{equation}
ds^2= \gamma_{\alpha \beta} dx^\alpha dx^\beta + h_{\mu\nu} dx^\mu
dx^\nu\;\;\; (1\leq \alpha , \beta \leq 2;\; 3\leq \mu ,\nu \leq n),
\end{equation}
where $\gamma_{\alpha \beta}$ depends on $\psi, r$ only, and
$h_{\mu \nu}r_i^{-2}$ is the metric tensor of the base manifold.
Inside the brane the energy-momentum tensor is
\begin{equation}
T^q_{\;\; m}= \epsilon_i (\psi, r)
\mbox{diag}(0,0,1,1,\cdots,1)\;\;\; (1 \leq q, m \leq n).
\end{equation}

The Einstein equation reads
\begin{equation}
R_{\alpha \beta} - \frac{1}{2} R \gamma_{\alpha \beta} = 0,
\end{equation}
\begin{equation}
R_{\mu \nu}- \frac{1}{2} R h_{\mu \nu} = - \epsilon_i (\psi, r)
h_{\mu \nu}.
\end{equation}
From above one can derive
\begin{equation}
R = 2\epsilon_i (\psi, r)
\end{equation}
and
\begin{equation}
R_{\mu \nu} = 0,\;\;\; R_{\alpha \beta} = \epsilon_i (\psi,
r)\gamma_{\alpha \beta }.
\end{equation}
That is in the scale of the brane thickness, the base manifold is
nearly Ricci flat, in the sense that the bulk cosmological
constant $\Lambda$ is much smaller than $\alpha_i$ and so omitted
here. It is noted that for the product spacetime, $R_{\alpha
\beta}$ remains intact for the reduction $\bar{M}_n
\longrightarrow \bar{M}_{2}$.

Substituting (18) and (12) into (11), and letting the thickness
approach zero, we see $\epsilon_i (\psi, r) \longrightarrow
\sigma_i \delta (r - r_i)/(\pi |r- r_i|)$, and
\begin{equation}
 \alpha_i = \sigma_i.
\end{equation}
This relation is a generalization of the result of string in
4-dimensional spacetime [11]. Apparently, the codimension of 2 is
critical for this argument. The higher co-dimension ``thick" brane
has been discussed in  [12], which are based on an axis symmetry.
Our argument is more general.

Now, for our instanton, the deficit angles or the tensions of the
two branes are determined by $\beta$
\begin{equation}
\sigma_i =  2\pi - \kappa_i|\beta| .
\end{equation}
Once the tension of one brane is given, both $\beta$ and the
tension of the other brane is fixed by eq. (20).

The action of the brane can be written as
\begin{equation}
I_{brane} =\frac{1}{8 \pi}(\sigma_l A_l + \sigma_k A_k).
\end{equation}

The total action is the sum
\begin{equation}
I = I_{gravity} + I_{brane} = -\frac{|\beta|}{8\pi} (\kappa_lA_l
+\kappa_k A_k).
\end{equation}

Now we consider the Kerr-like instanton [10][13]
\[
ds^2 = \frac{\Delta_r}{\rho^2} \left [ d\chi + \frac{\alpha}{\Xi}
\sin^2 \theta d\phi \right ]^2 + \frac{\rho^2}{\Delta_r} dr^2
+\frac{\rho^2}{\Delta_\theta} d\theta^2 +
\frac{\Delta_\theta}{\rho^2} \sin^2\theta \left [ \alpha d\chi -
\frac{r^2 - \alpha^2}{\Xi} d\phi \right ]^2
\]
\begin{equation}
+ r^2 \cos^2\theta d \Omega^2_{n-4},
\end{equation}
where $\alpha$ is the angular momentum parameter for the imaginary time
$\chi$, $d\Omega^2_{n-4}$ is a general unit elliptic space, and
\[
\Delta _r = (r^2 -\alpha^2) \left ( 1 - \frac{2 \Lambda
r^2}{(n-1)(n-2)} \right )  - \frac{2m}{r^{n-5}},
\]
\[
\Delta_\theta = 1 - \frac{2\Lambda
\alpha^2}{(n-1)(n-2)}\cos^2\theta,
\]
\[
\Xi = 1 -\frac{2\Lambda \alpha^2}{(n-1)(n-2)},
\]
\begin{equation}
\rho^2 = r^2 - \alpha^2 \cos^2 \theta.
\end{equation}

The Killing coordinate $\chi$ is identified by a period $\beta$ as
in the black hole creation scenario [14], i.e one can set $0 \leq
\chi \leq \beta |\Xi|^{-1}$. The instanton is constructed from a
sector between two neighboring horizons $r_l \leq r \leq r_k$,
with $\Delta_r \geq 0$. In general, at least one conical
singularity arises. Again, these conical singularities are
identified as the branes. The Einstein equation should apply to
these singularities (branes) as well. By the same argument as for
the nonrotating case, the brane tension should equal the deficit
angle as in (19)(20).

The double continuation can be carried out through an analytic
continuation from $d\Omega^2_{n-4}$ into a de Sitter-like
spacetime.

Using the same method, the gravitational action  in (1) can be derived
\begin{equation}
I_{gravity} = - \frac{1}{4} (A_k + A_l),
\end{equation}
where the horizon area is
\begin{equation}
A_i = \left |\frac{4\pi(r^{n-2}_i - \alpha^2
r^{n-4}_i)\Omega_{n-4}}{(n-3)\Xi}\right |
\end{equation}
and the surface gravity is
\begin{equation}
\kappa_i = \left | \frac{1}{2\Xi (r^2_i - \alpha^2)} \frac{d\Delta_r}{dr}\right |_{r =r_i}.
\end{equation}

The action of the brane in (1) can be written in the same form as (21).
The total action takes the same form (22).

The relative creation probability, at the $WKB$ level, is
\begin{equation}
P \approx  exp (-I).
\end{equation}

In the probability calculation of the black hole creation, the
induced metric and the matter  content on the equator of the
constrained instanton are given. These constraints can be
characterized by a few parameters, like mass $m$, charge $Q$ and
angular momentum $J$. Therefore, the path integral can be
interpreted as the partition function $Z$ for a microcanonical
ensemble in gravitational thermodynamics [14]. In this ensemble,
the entropy is
\begin{equation}
S = \ln Z \approx -I.
\end{equation}
The second equality is due to the $WKB$ approximation, in which
the path integral is evaluated by the instanton contribution. The
neglected contribution is from the fluctuation around the
background. There is no contribution from the brane tension.

However, in the braneworld there are two parts in the entropy. One
part is due to the tension of the brane. This is new. One may
wonder why this phenomenon was not considered in the black hole
case, say the Schwarzschild-de Sitter case. The reason is that the
surfaces of quantum transitions, or the equators for braneworld
and black hole creations are different. In the latter case, the
3-metric of the equator is the only configuration for the wave
function there, while in the former case, the brane tensions (or
the deficit angles) must be included in the configuration. The
right choice of the representation is also crucial for
dimensionality in quantum cosmology [15].

The other contribution to the entropy is due to the gravity, as in
the black hole with a 4-dimensional de Sitter background. The
result on entropy for the nonrotating higher-dimensional black
hole with distinct surface gravities is not new [14]. It is noted that for the black
hole entropy one has to replace $\alpha$ by $ia$ in the above
formulas, where $a$ is the angular momentum parameter for the real Killing
time. In the braneworld the observer is living in a brane. Let us
take the nonrotating case with $k =1$ as an example, the brane is
a de Sitter spacetime, and the observer perceives his world only
through non-gravitational forces, then he would conclude that the
universe is created from a $n-2-$dimensional instanton, i.e.
$S_{n-2}$ sphere. In the traditional cosmology, the entropy is one
quarter of area of the $n-4-$dimensional horizon $S_{n-4}$. In the
braneworld the entropy associated with one brane is one quarter of
the volume of $S_{n-2}$. In the new scenario, both the two branes
and associated tensions contribute to the entropy of the universe.

If one takes the braneworld seriously, one has to evaluate the
entropy of the universe in the new way. Since at least one kind of
interaction penetrates into the bulk, the $n-2-$dimensional brane
is not self-contained. This problem is hidden in the traditional
Kaluza-Klein models, since the effect of the extra dimensions to
the entropy and the action has been automatically taken into
account in redefining the Newton gravitational constant.

From (22) and (29) it follows that the entropy of those
braneworlds can take arbitrarily large values as the parameter
$|\beta|$, or the tensions, increase. This result of calculation
is not only interesting for the braneworld scenario, but also for
quantum gravity in general.

There is a new perspective on the origin of the cosmological constant, the so-called "$\Lambda-N$ correspondence" [16]. It is conjectured that the cosmological constant $\Lambda$ is a direct consequence of the finite number of states $e^N$ in the Hilbert space describing the world. Many people believe that, in any universe with a positive $\Lambda$
the observable entropy $S$ is less than or equal to $N$. For a given $\Lambda$, $N$ is saturated by the entropy of de Sitter space, and one has
\begin{equation}
N = \left [\frac{(n-1)(n-2)}{2\Lambda}\right ]^{\frac{n-2}{2}}\frac{\Omega_{n-2}}{4}.
\end{equation}
Bousso is able to show that this is true for all spherical symmetric
universes with $n \geq 4$ [16]. He suggested that $\Lambda >0$ may be a sufficient condition for the entropy bound $S \leq N$ without the symmetry condition [17].

However, for $n>4$ case, Bousso, DeWolfe and Myers realized that the conjecture
does not hold by providing a counterexample, which is a product space with
flux of the form $(A)dS_p \times S^q$ [18].

The model provided in this letter violates
the ``$N$-bound" conjecture not only for $n>4$, but also for $n=4$. Since our model is not spherically symmetric, this counterexample does not conflict with Bousso's proof for the spherically symmetric 4-dimensional model [16]. Therefore, it is concluded that, even for $n=4$, the specification of a positive $\Lambda$ is not sufficient to characterize the class of spacetime described by quantum gravity theories with finite-dimensional Hilbert space.

\vspace*{0.3in} \rm

\bf Acknowledgement:

\vspace*{0.1in} \rm

I would like to thank Xin Xin Du and An Zhong Wang for discussions.

\vspace*{0.3in} \rm

\bf References:

\vspace*{0.1in} \rm

1. N. Arkani-Hamed, S. Dimopoulos and G. Dvali, \it Phys. Lett.
\rm \bf B\rm\underline{429}, 263 (1998); I. Antoniadis, N.
Arkini-Hamed, S. Dimopoulos and G. Dvali, \it Phys. Lett. \rm \bf
B\rm\underline{436}, 257 (1998); N. Arkani-Hamed, S. Dimopoulos
and G. Dvali, \it Phys. Rev. \rm \bf D\rm\underline{59}, 086004
(1999).

2. L. Randall and R. Sundrum, \it Phys. Rev. Lett. \rm
\underline{83}, 3370 (1999); L. Randall and R. Sundrum, \it Phys.
Rev. Lett. \rm \underline{83}, 4690 (1999).

3. P. Horava and E. Witten, \it Nucl. Phys. \rm \bf
B\rm\underline{460}, 506 (1996); P. Horava and E. Witten, \it
Nucl. Phys. \rm \bf B\rm\underline{475}, 94 (1996).

4. J.B. Hartle and S.W. Hawking, \it Phys. Rev. \rm \bf D\rm
\underline{28}, 2960 (1983).

5. Z.C. Wu, \it Phys. Rev. \rm \bf D\rm \underline{31}, 3079
(1985); S.W. Hawking and N. Turok, \it Phys. Lett. \rm \bf B\rm
\underline{425}, 25 (1998);  N. Turok and S.W. Hawking, \it Phys.
Lett. \rm \bf B\rm \underline{432}, 271 (1998).

6.W. Hawking, T. Hertog and H.S. Reall, \it Phys. Rev. \rm \bf
D\rm\underline{62}, 043501 (2000); K. Koyama and J. Soda, \it
Phys. Lett. \rm \bf B\rm\underline{483}, 432 (2000).

7. J.W. Chen, M.A. Luty and E. Ponton, \it JHEP
\rm\underline{0009}, 012 (2000); J.M. Cline, J. Descheneau, M.
Giovannini and J. Vinet, \it JHEP \rm\underline{0306}, 048 (2003);
P. Bostock, R. Gregory, I. Navarro and J. Santiago, hep-th/0311074;
R.A. Battye, B. Carter and A. Mennim, hep-th/0312198;  H.M. Lee
and G. Tasinato, hep-th/0401221; I. Navarro and J. Santiago,
hep-th/0402204.

8. E. Witten, \it Nucl. Phys. \rm\bf B\rm\underline{195}, 481
(1982).

9. O. Aharony, M. Fabinger, G.T. Horowitz and E. Silverstein, \it
JHEP \rm\underline{0207}, 007 (2002); F. Dowker, J.P. Gauntlett,
G.W. Gibbons and G.T. Horowitz, \it Phys. Rev. \rm \bf
D\rm\underline{52}, 6929 (1995); F. Dowker, J.P. Gauntlett, G.W.
Gibbons and G.T. Horowitz, \it Phys. Rev. \rm \bf
D\rm\underline{53}, 7115 (1996).

10. D. Birmingham and M. Rinaldi, \it Phys. Lett. \rm \bf B\rm
\underline{544}, 316 (2002); B. Balasubramanian and S.F. Ross, \it
Phys. Rev. \rm\bf D\rm\underline{66}, 086002 (2002).

11. See for example, A. Vilenkin and E.P.S. Shellard, \it Cosmic
Strings and Other Topological Defects, \rm Cambridge University
Press (1994).

12. R. Gregory, \it Phys. Rev. Lett. \rm \bf \rm\underline{84},
2564 (2000); I. Olasagasti and A. Vilenkin, \it Phys. Rev. \rm \bf
D\rm\underline{62}, 044014 (2000); M. Giovannini, H. Meyer and M.
Shaposhnikov, \it Nucl. Phys. \rm \bf B\rm\underline{619}, 615
(2001); S. Kanno and J. Soda, hep-th/0404207.

13. S.W. Hawking, C.J. Hunter and M.M. Taylor-Robinson, \it Phys.
Rev. \rm \bf D\rm\underline{59}, 064005 (1999); D. Klemm, V.
Moretti and L. Vanzo, \it Phys. Rev. \rm \bf D\rm\underline{57},
6127 (1998), Erratum-ibid. \rm \bf D\rm\underline{60}, 109902
(1999); D. Klemm, \it JHEP \rm\underline{9811}, \rm 019 (1998); M.H.
Dehghani, \it Phys. Rev. \rm \bf D\rm\underline{65}, 124002 (2002).

14. Z.C. Wu, \it Int. J. Mod. Phys. \rm \bf D\rm\underline{6}, 199
(1997); Z.C. Wu, \it Int. J. Mod. Phys. \rm \bf D\rm\underline{7},
111 (1998); R. Bousso and S.W. Hawking, \it Phys. Rev. \rm \bf
D\rm \underline{59}, 103501 (1999), Erratum-ibid. \rm \bf D\rm
\underline{60}, 109903 (1999); Z.C. Wu, \it Phys. Lett. \rm \bf
B\rm \underline{445}, 274 (1999); Z.C. Wu, \it Gene. Relativ.
Grav. \rm\underline{32}, 1823 (2000).

15. Z.C. Wu, \it Gene. Relativ. Grav. \rm\underline{34}, 1121
(2002); Z.C. Wu, \it Phys. Lett. \rm\bf B\rm\underline{585}, 6
(2004).

16. W. Fischler, unpublished; W. Fischler, \it Taking de Sitter seriously. \rm Talk given at \it Role of Scaling Laws in Physics and Biology (Celebrating the 60th Birthday of Geoffrey West), \rm Santa Fe, Dec. 2000; T. Banks, hep-th/0007146; R. Bousso, \it JHEP \rm\underline{0011}, \rm 038 (2000).

17. R. Bousso, \it Rev. Mod. Phys. \rm\underline{74}, \rm 825 (2002).

18. R. Bousso, O. DeWolfe and R.C. Myers, \it Found. Phys. \rm\underline{33}, \rm 297 (2003).

\end{document}